\documentclass[a4paper,10pt,twoside]{cpc-hepnp}
\usepackage{CJK,upgreek,fancyhdr}
\usepackage{multicol}
\usepackage{graphicx}
\usepackage{booktabs}
\usepackage{amssymb,bm,mathrsfs,bbm,amscd}
\usepackage[tbtags]{amsmath}
\usepackage{lastpage}
\usepackage{color}

\begin{document}
\begin{CJK*}{GB}{gbsn}

\fancyhead[c]{\small Chinese Physics C~~~Vol. xx, No. x (201x) xxxxxx}
\fancyfoot[C]{\small 010201-\thepage}

\footnotetext[0]{Received \today}

\title{Pairing-energy coefficients of neutron-rich fragments in spallation reactions \thanks{Supported by the National Natural Science Foundation of China (U1732135), Natural Science Foundation of Henan Province (162300410179), and Henan Normal University for the Excellent Youth (154100510007).}}

\author{%
\quad Fei Niu $^{1,2}$%
\quad Chun-Wang Ma$^{1}$ $^{*}$\email{machunwang@126.com}
}
\maketitle

\address{%
$^1$ Institute of Particle and Nuclear Physics, Henan Normal University, Xinxiang 453007, People's Republic of China\\
$^2$ Institute of Modern Physics, Chinese Academy of Sciences, Lanzhou, 730000, People's Republic of China\\
}

\begin{abstract}
The ratio of pairing-energy coefficient to temperature ($a_{p}/T$) of neutron-rich fragments produced in spallation reactions has been investigated by adopting an isobaric yield ratio method deduced in the framework of a modified Fisher model. A series of spallation reactions, 0.5$A$ and 1$A$ GeV $^{208}$Pb + $p$, 1$A$ GeV $^{238}$U + $p$, 0.5$A$ GeV $^{136}$Xe + $d$, 0.2$A$, 0.5$A$ and 1$A$ GeV $^{136}$Xe + $p$, and $^{56}$Fe + $p$ with incident energy ranging from 0.3$A$ to 1.5$A$ GeV, has been analysed. An obvious odd-even staggering is shown in the fragments with small neutron excess ($I\equiv N - Z$), and in the relatively small-$A$ fragments which have large $I$. The values of $a_{p}/T$ for the fragments, with $I$ from 0 to 36, have been found to be in a range from -4 to 4, and most values of $a_{p}/T$ fall in the range from -1 to 1. It is suggested that a small pairing-energy coefficient should be considered in predicting the cross sections of fragments in spallation reactions. It is also concluded that the method proposed in this article is not good for fragments with $A/A_{s} >$ 85\% (where $A_{s}$ is the mass number of the spallation system).
\end{abstract}


\footnotetext[0]{\hspace*{-3mm}\raisebox{0.3ex}{$\scriptstyle\copyright$}2013
Chinese Physical Society and the Institute of High Energy Physics
of the Chinese Academy of Sciences and the Institute
of Modern Physics of the Chinese Academy of Sciences and IOP Publishing Ltd}%

\begin{multicols}{2}

\section{Introduction}
The pairing energy, which originates from neutron and proton pairing in the nucleus, contributes to the binding energy of a nucleus. The study of  nucleon (proton/neutron) correlations and nuclear structure in neutron-rich nuclei has become a hot topic, because their pairing correlations are quite different to those of symmetric nuclei.
The Facility for Rare Isotope Beams (FRIB) under construction at Michigan State University, and the suggested Beijing Isotope Online Separator Laboratory (BISOL) aiming at the study of nuclei with large asymmetries, will provide new opportunities to study neutron-rich isotopes near the drip line. Meanwhile, the design of accelerator-driven systems (ADS) requires a precise knowledge of the production cross sections inside the spallation target. These new opportunities will surely  motivate research into very asymmetric nuclear matter, and test nuclear theories in extreme situations.

The pairing energy, which can be divided into isoscalar and isovector parts, also depends on temperature ($T$). At finite temperature, for example in the fragments produced in reactions, the pairing energy is weakened and even disappears \cite{MRH10PRC,Cw11,Cw16}, because of the disappearance of the delicate balance between isoscalar and isovector pairing energy \cite{Kkprc05}. Parameterizations for $T$-dependent binding energy for finite nuclei have been suggested by Lee and Mekjian \cite{ParaTBE} using a density-functional theory based on a Skyrme interaction. The binding energies of nuclei at finite temperature have also been investigated, using the relativistic Hartree-Bogoliubov theory \cite{NYF13PRC,Lisboa16PRC,WNiu17CPC}, covariant density functional theory \cite{CDFT}, Hartree-Fock-BCS approximation \cite{HFBCS}, etc. Formulas for $T$-dependent binding energy are usually used to calculate the excitation energy (or free energy) of the primary fragment, such as in the statistical multifragmentation model \cite{Rdcpc01} and the thermodynamic model \cite{Cbpr05}. Though the pairing correlation of nucleons becomes less important when the temperature goes high, it is very important in forming cold fragments. The yields of fragments are significantly influenced by the pairing correlations. The pairing correlations also take important roles in the simulation of sequential decay \cite{FragT12Ma,FragT13Ma}. The temperature effects in binding energy are usually omitted. Examples can be found in the de-excitation calculation of the statistical abrasion-ablation model \cite{ma14epja} and the simulated annealing clusterization algorithm (SACA) method \cite{SACA1,SACA2,SACA3,SACA4}. In a previous analysis of the pairing energies of fragments in projectile fragmentation reactions, it is suggested that the pairing-energy coefficients are rather low compared to the standard value \cite{MRH10PRC,Cw11,Cw16}. This phenomenon is explained as a temperature effect in the pairing energy by a self-consistent finite-temperature relativistic Hartree-Bogoliubov model \cite{NYF13PRC}.

The yields of fragments are mostly determined by the free energy of fragments in theories, for example the thermodynamical models such as the canonical ensemble theories \cite{GCE01,GCE07,Cbpr05}, the statistical models such as the modified Fisher model \cite{MFM82,MFM84}, and the Landau free energy theory \cite{MRH10PRC}, etc. A set of parameterizations, which is called the semi-empirical parameterizations of SPAllation residue Cross Section ({\sc spacs}) \cite{SPACS-code}, with the physical ideas originating from {\sc epax}, predicts the cross sections of fragments in spallation reactions well. It is also interesting to find that the isobaric difference between fragments of different neutron-richness ($I\equiv N-Z$) in spallation reactions shows a scaling phenomenon like those in the projectile fragmentation reactions \cite{Ma17PRCIBD,Ma17JPG}.  Many theories can deal with both projectile fragmentation and spallation reactions \cite{SpalBChM}. It is generally accepted that in dynamical models, global equilibrium of the system is hard to achieve (though local equilibrium can be achieved), though in thermodynamics the equilibrium of the system is a fundamental assumption. The fragments measured in experiments, both in spallation reactions and projectile fragmentation reactions, undergo a sequential decay process. The study of fragments produced in projectile fragmentation reactions should help to understand the fragment production in spallation reactions. The pairing-energy coefficients of neutron-rich fragments in projectile fragmentation reactions have been analyzed using an isobaric yield ratio (IYR) method. It is proposed that many terms can be cancelled out in the isobaric methods \cite{Cw16,Cw11,MRH10PRC,Ma17PRCIBD,Cwprc15,Tasym16Q,Tasym16D,NST2015IBD,Ma15PLB,Ma16JPG}. In this work, more neutron-rich fragments produced in spallation reactions will be analyzed to understand the pairing energy of very neutron-rich fragments.

The article is organized as following. In Section 2, 
the isobaric yield ratio method to determine the pairing energy coefficient  is briefly introduced.
In Section 3, 
a series of spallation reactions is analyzed, and the results for the pairing-energy coefficient of neutron-rich fragments are presented and discussed.
In Section 4, 
conclusions are presented.

\section{Formulism}
\label{fmlsm}
The isobaric yield ratio method deduced in the framework of a modified Fisher model (MFM) \cite{MFM82,MFM84,MRH10PRC,Cw11}, as has been discussed in detail in Ref. \cite{Cw16}, will be adopted to analyze the pairing-energy-coefficient of the very neutron-rich fragments in  spallation reactions. In this article, the method will be briefly introduced. The MFM is a thermodynamics model, which assumes that the system is in equilibrium. In the MFM, the cross section ($\sigma$) of a fragment is mainly determined by the free energy, temperature, and chemical potentials of protons and neutrons, and has the form \cite{MFM82,MFM84}
\begin{equation}\label{crsmfm}
\begin{split}
\sigma(I,A)=&CA^{-\tau}exp \{[W(I,A) + \mu_{n}N + \mu_{p}Z]/T \\
& + N\ln(N/A) + Z\ln(Z/A)\},
\end{split}
\end{equation}
where $I = N - Z$ and $A$ are the neutron excess and the mass number of the fragment, respectively. $\tau$ depends on the asymmetry of the reaction system \cite{Mh11}. $\mu_{n} (\mu_{p})$ is the chemical potential of the neutrons (protons). $W(A,I)$ is the free energy of the fragment, which reads as a form of the Weisz\"{a}cker-Bethe semiclassical mass formula \cite{Mh10,Cf35,Ha36}
\begin{equation}\label{wenergy}
\begin{split}
W(I,A)=&a_{v}(\rho,T)-a_{s}(\rho,T)A^{2/3} \\
& -a_{sym}(\rho,T)I^{2}/A-E_{c}-\delta(I,A),
\end{split}
\end{equation}
where $a_{v}$, $a_{s}$, and $a_{sym}$ represent the coefficients of the volume, surface, and symmetry energies, respectively. The dependence of $W(I,A)$ on the density $\rho$ and temperature $T$ of the system is reflected by $a_{i}(\rho,T)$. $E_{c}$ is the Coulomb energy. The following pairing energy $\delta(I,A)$ is adopted \cite{Ae53}:
\begin{eqnarray}
\delta(I,A)=\left\{
\begin{aligned}
&a_{p}(\rho,T)/A^{1/2},  &(o-o)\\
&0,  &(o-e)\\
&-a_{p}(\rho,T)/A^{1/2},  &(e-e)
\end{aligned}
\right.
\end{eqnarray}
where $a_{p}$ is the pairing-energy coefficient. $a_{i}$ can be seen as a free parameter and can be fitted from the fragment yield \cite{MFM82,MFM84,Cw11}. Different forms of pairing energy can be found in Refs. \cite{Cw16,SnS2n17CPC,pairing01,pairing02}.

From Eq. (\ref{crsmfm}), the IYR between isobars differing 2 units in $I$ is defined as
\begin{equation}\label{deflnr}
\begin{split}
\ln R(I+2,I,A)=& \ln[\sigma(I+2,A)/\sigma(I,A)] \\
=& [W(I+2,A)-W(I,A)+(\mu_{n}-\mu_{p})]/T \\
& +S_{mix}(I+2,A)-S_{mix}(I,A),
\end{split}
\end{equation}
where $S_{mix}(I,A) = N\ln(N/A) + Z\ln(Z/A)$. The dependence of $\sigma$ on $C$ and $\tau$ in Eq. (\ref{crsmfm}) cancels out for isobars. Inserting Eq. (\ref{wenergy}) into Eq.~(\ref{deflnr}), one obtains
\begin{equation}\label{lnravs}
\begin{split}
\ln R(I+2,I,A)=& [(\mu_{n}-\mu_{p})-4a_{sym}(I+1)/A \\
&+\Delta E_{c}(I+2,I,A)+\Delta \delta(I+2,I,A)]/T \\
&+\Delta S_{mix}(I+2,A),
\end{split}
\end{equation}
in which $\Delta E_{C}(I + 2, I, A)$ is the difference between the Coulomb energies of isobars, and $\Delta S_{mix}(I + 2, I, A) = S_{mix}(I + 2, A) - S_{mix}(I, A)$. $\Delta\delta(I + 2, I, A)$ is the difference between the pairing energies of the isobars with $(I + 2, A)$ and $(I, A)$. In Eq. (\ref{lnravs}), the volume energies, at the same time as the surface energies of isobars, cancel out. For a fragment with even $I$, $a_{p}/T$ can be written as
\begin{equation}\label{aptcal}
\begin{split}
(sgn)a_{p}/T=& (1/2)A^{1/2}\{\ln R(I+2,I,A)-[(\mu_{n}-\mu_{p}) \\
& +2a_{c}(Z-1)/A^{1/3}-4a_{sym}(I+1)/A]/T \\
& -\Delta S_{mix}(I+2,A)\},
\end{split}
\end{equation}
where $sgn=1$ and $sgn=-1$ denote the odd-odd and even-even fragments, respectively. In Eq. (\ref{aptcal}), the terms  relating to the chemical potential and the symmetry energy are replaced by the neighboring IYRs with odd-$I$, i.e., [$\ln R(I + 3,I + 1,A^{'})$ and $\ln R(I + 1, I - 1, A^{'})$] which have no pairing energy. For a neutron-rich fragment ($I >$ 0), one has \cite{Cw16}
\begin{equation}\label{apcal}
\begin{split}
(sgn)&a_{p}/T= \\
& (1/2)A^{1/2}\{\ln R(I+2,I,A)-(1/2) \\
& \times[\ln R(I+3,I+1,A^{'})+\ln R(I+1,I-1,A^{'})] \\
& +(1/2)[\Delta S_{mix}(I+3,I+1,A^{'}) \\
& +\Delta S_{mix}(I+1,I-1,A^{'})]-\Delta S_{mix}(I+2,I,A)\},
\end{split}
\end{equation}
where $A^{'}$ denotes that for the odd-$I$ fragment, the mass number differs 1 unit from that of the even-$I$ fragment. For convenience, $<IYR> = \ln R(I+3,I+1,A^{'}) + \ln R(I+1,I-1,A^{'})$ is defined.

\begin{center}
\includegraphics[width=8.cm]{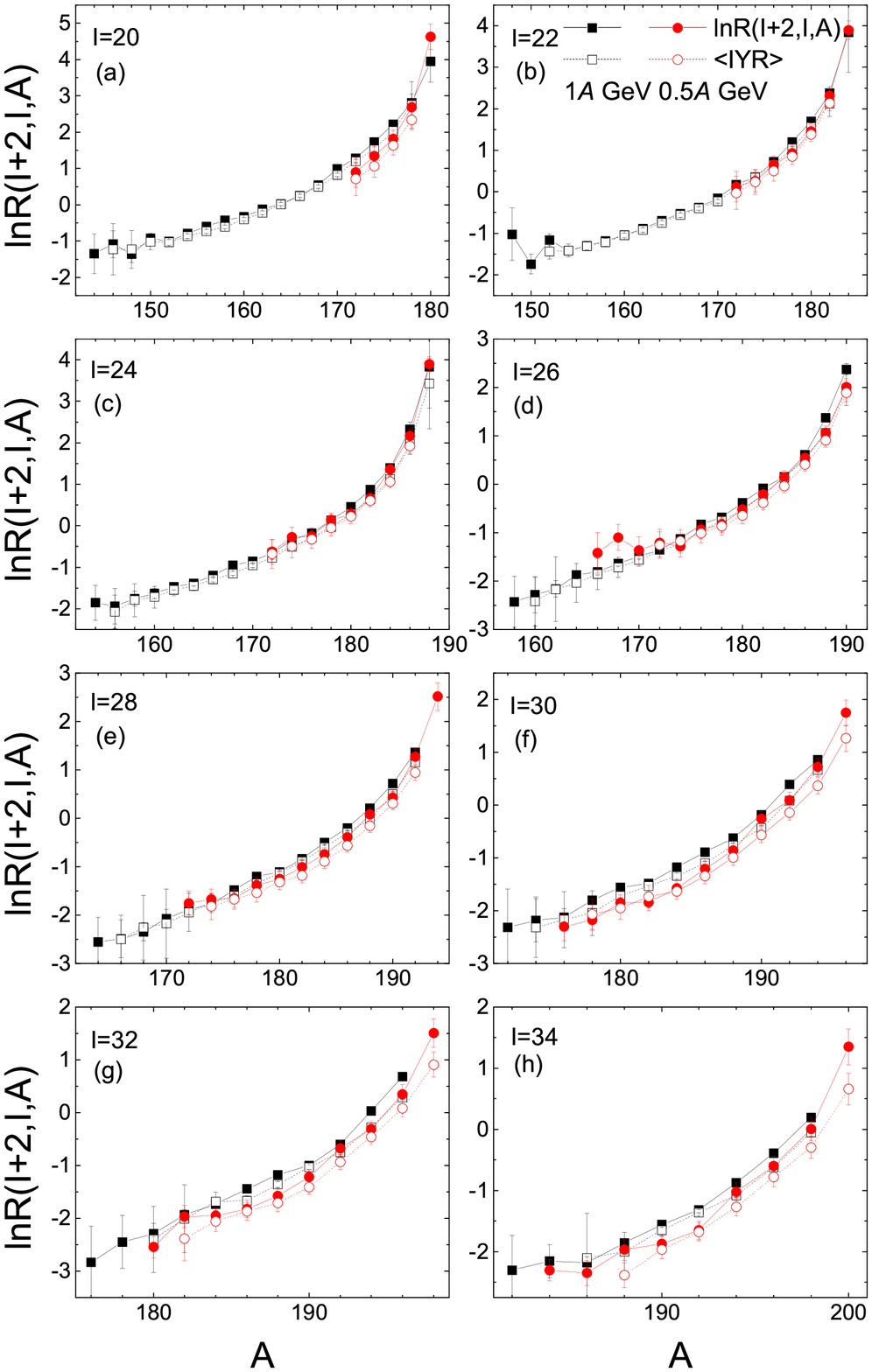}
\figcaption{\label{Pbp51000}  $\ln R(I+2,I,A)$ (solid symbols) and $<IYR>$ (open symbols) for the measured fragments in the 0.5$A$ and 1$A$ GeV $^{208}$Pb + $p$ spallation reactions. The measured cross sections are taken from Refs. \cite{Pb208p500} and \cite{Pb208p1000}, respectively.}
\end{center}

\section{Results and discussion}
\label{results}
It is important to verify whether $<IYR>$ is a good approximation for the symmetry energy and chemical potential energy terms in Eq. (\ref{apcal}) for the more neutron-rich fragments in spallation reactions, though this has already been done for fragmentation reactions \cite{Ma17PRCIBD}. The spallation reactions 0.5$A$ GeV $^{208}$Pb + $p$  have been measured by L. Audouin \textit{et al} using the inverse-kinematics method and the fragment separator (FRS) spectrometer at GSI \cite{Pb208p500}. The spallation reaction $^{208}$Pb + $p$ has been measured at a higher energy of 1$A$ GeV by T. Enqvist \textit{et al} using the FRS at GSI, and the isotopes from $Z =$ 22 to 82 have been identified \cite{Pb208p1000}. The results of $\ln R(I+2,I,A)$ and $<IYR>$ for the fragments produced in the 0.5$A$ and 1$A$ GeV $^{208}$Pb + $p$ spallation reactions, with $I$ ranging from 20 to 34, are plotted in Fig. \ref{Pbp51000}. The distributions of $\ln R(I + 2,I,A)$ in the 0.5$A$ and 1$A$ GeV $^{208}$Pb + $p$ spallation reactions are similar. The incident energy only shows a small influence on $\ln R(I + 2,I,A)$ for the more neutron-rich fragments of $I \geq$ 30. The $\ln R(I + 2,I,A)$ for each $I$ chain shows a small odd-even staggering, while the staggering is weakened when $A$ becomes larger. Only small differences between $\ln R(I+2,I,A)$ and $<IYR>$ are seen for the fragments in the 1$A$ GeV $^{208}$Pb + $p$ reaction for fragments of $I \leq$ 26, while this difference becomes larger when $A$ becomes larger. A larger difference between $\ln R(I+2,I,A)$ and $<IYR>$ is seen for the more neutron-rich fragments with $I >$ 26. Similar behavior can be found in the 0.5$A$ GeV reaction, though the mass ranges of the measured fragments in the 0.5$A$ GeV reaction are much smaller than those in the 1$A$ GeV reactions when $I\leq$ 28, because the 1$A$ GeV reaction is much more violent than the 0.5$A$ GeV reaction. According to Eq. (\ref{apcal}), $a_p/T$ is mainly determined by the difference between $\ln R(I+2,I,A)$ and $<IYR>$, since the mixing terms of $N$ and $Z$ are negligible.

\begin{center}
\includegraphics[width=6.5cm]{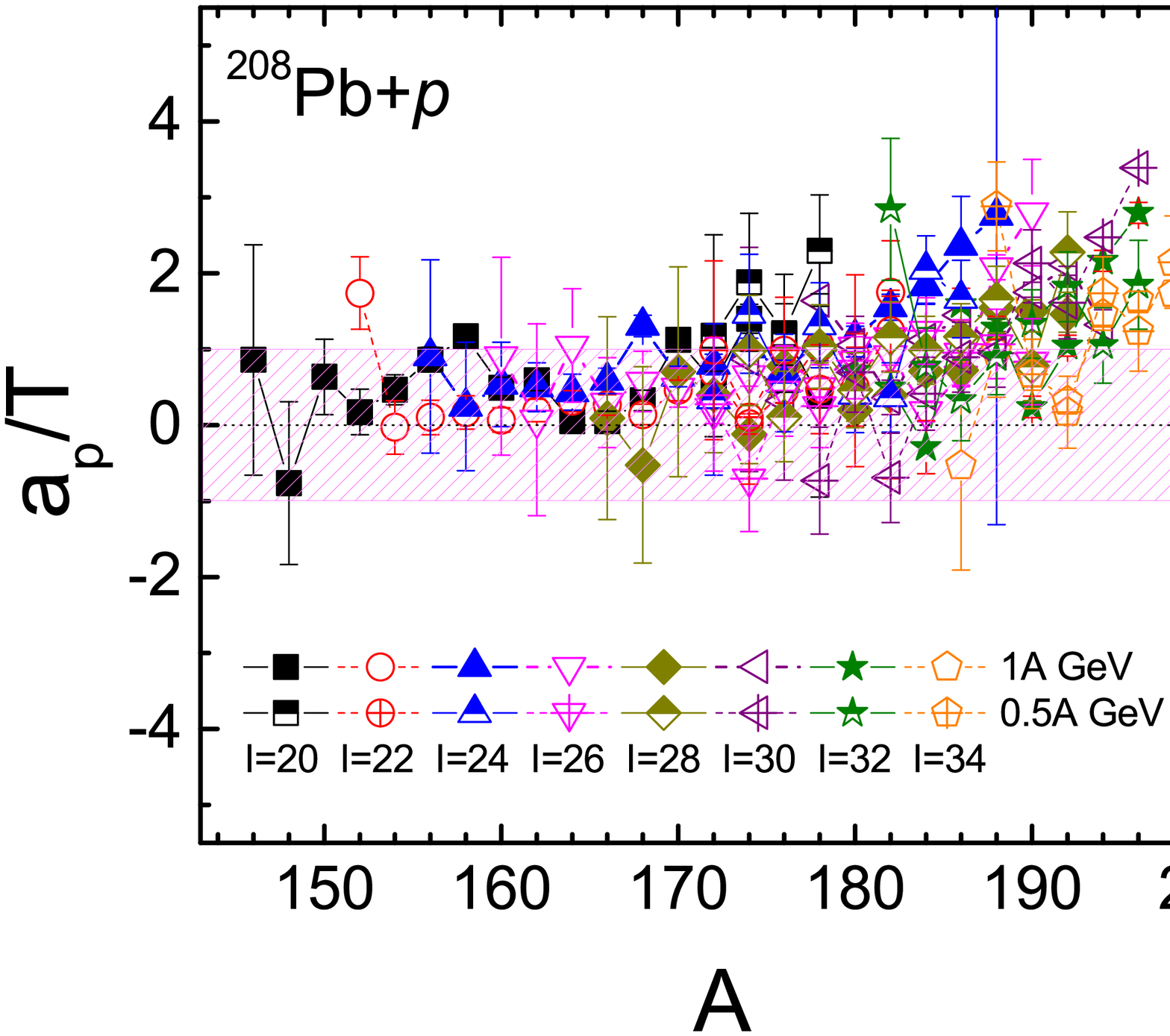}
\figcaption{\label{apPbp51000}  The values of $a_{p}/T$ determined by Eq. (\ref{apcal}) for the measured fragments in the 0.5$A$ and 1$A$ GeV $^{208}$Pb + $p$ spallation reactions. The shaded area denotes the range from -1 to 1.}
\end{center}

The results of $a_{p}/T$ determined by Eq. (\ref{apcal}) for the measured fragments in the 0.5$A$ and 1$A$ GeV $^{208}$Pb + $p$ spallation reactions are plotted in Fig. \ref{apPbp51000}. The values of $a_{p}/T$ for most of the fragments are within the range -1 to 1, and for many fragments the values of $a_{p}/T$ are near 0. For fragments with a relatively large $A$, $a_{p}/T$ increases. This can be accounted for by the large difference between $\ln R(I+2,I,A)$ and $<IYR>$ shown in Fig. \ref{Pbp51000}. The results of $a_{p}/T$ for most of the fragments in the 0.5$A$ GeV $^{208}$Pb + $p$ spallation reaction are also within the range -1 to 1. In both reactions, the results of $a_{p}/T$ in general do not exceed the range -2 to 4. The values of $a_{p}/T$ for fragments are found to increase with $A$ when $A >$ 180, which should be produced in peripheral reactions.

\begin{center}
\includegraphics[width=8cm]{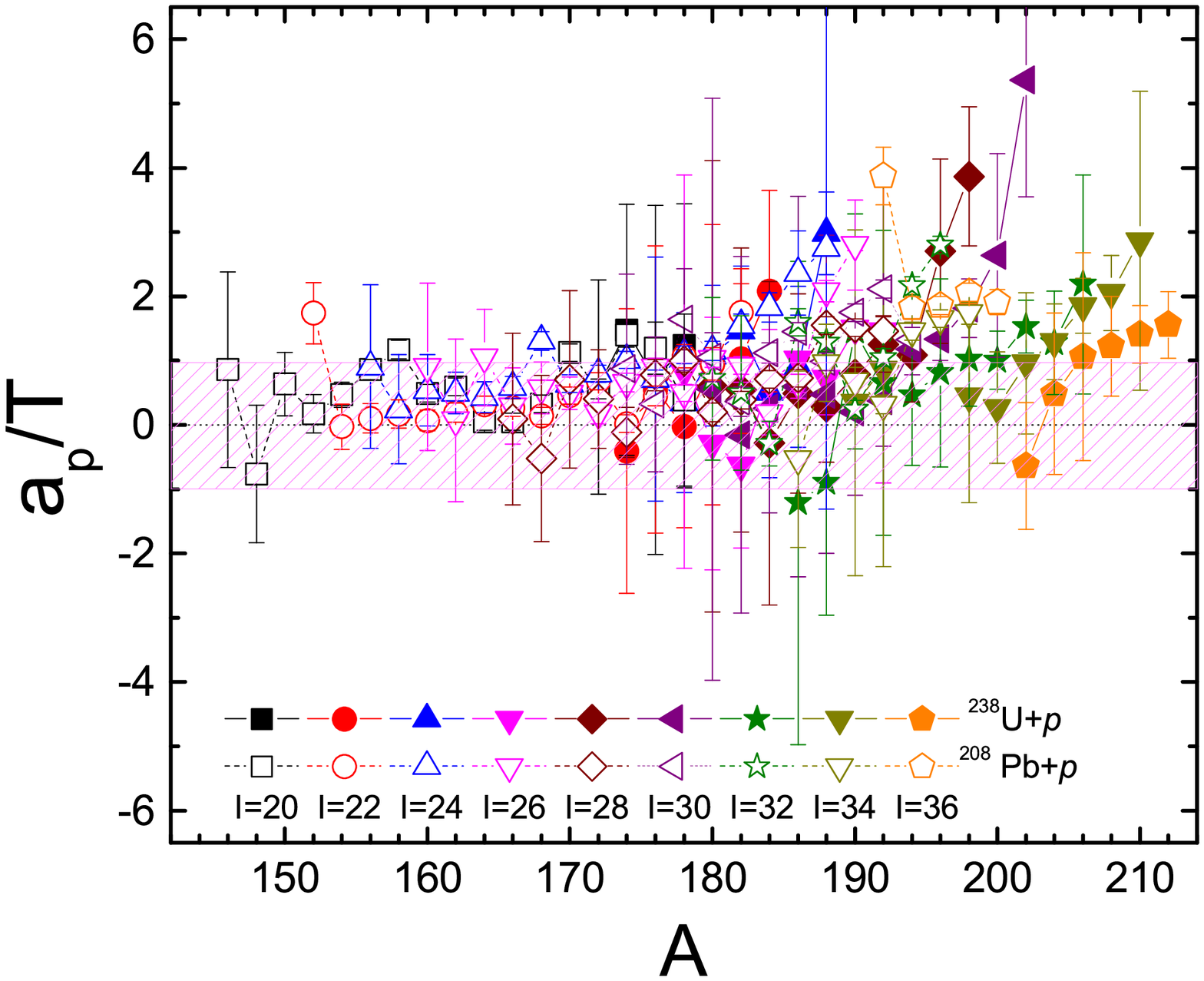}
\figcaption{\label{U238Pb208p}  A comparison of $a_{p}/T$ for the measured fragments in 1$A$ GeV $^{238}$U + $p $ and $^{208}$Pb + $p$ spallation reactions. The cross sections of fragments measured in the $^{238}$U and $^{208}$Pb reactions are taken from Refs. \cite{U239p} and \cite{Pb208p1000}, respectively. The shaded area denotes the range from -1 to 1.}
\end{center}

The cross sections of fragments produced in 1$A$ GeV $^{238}$U + $p$ spallation reactions have been measured by J. Ta\"{\i}eb \textit{et al} using the FRS at GSI \cite{U239p}. The cross sections of more than 350 isotopes from tungsten ($Z =$ 74) to uranium ($Z =$ 92) have been measured using the inverse kinematics method. The values of $a_{p}/T$ for the fragments in the 1$A$ GeV $^{238}$U + $p$ and $^{208}$Pb + $p$ (the same as in Fig. \ref{apPbp51000}) spallation reactions are compared in Fig. \ref{U238Pb208p} to observe the evolution of $a_{p}/T$ in neutron-rich fragments. For the fragments of $A <$ 180, the values of $a_{p}/T$ for the fragment in the $^{238}$U reaction are similar to those in the $^{208}$Pb reaction. One can see that $a_{p}/T$ for most of the fragments in the $^{238}$U + $p$ reaction are also within the range from -1 to 1, and $a_{p}/T$ is found, in general, to increase with $A$. However, the values of $a_{p}/T$ do not exceed the range from -2 to 4, except for a few fragments.

\begin{center}
\includegraphics[width=8cm]{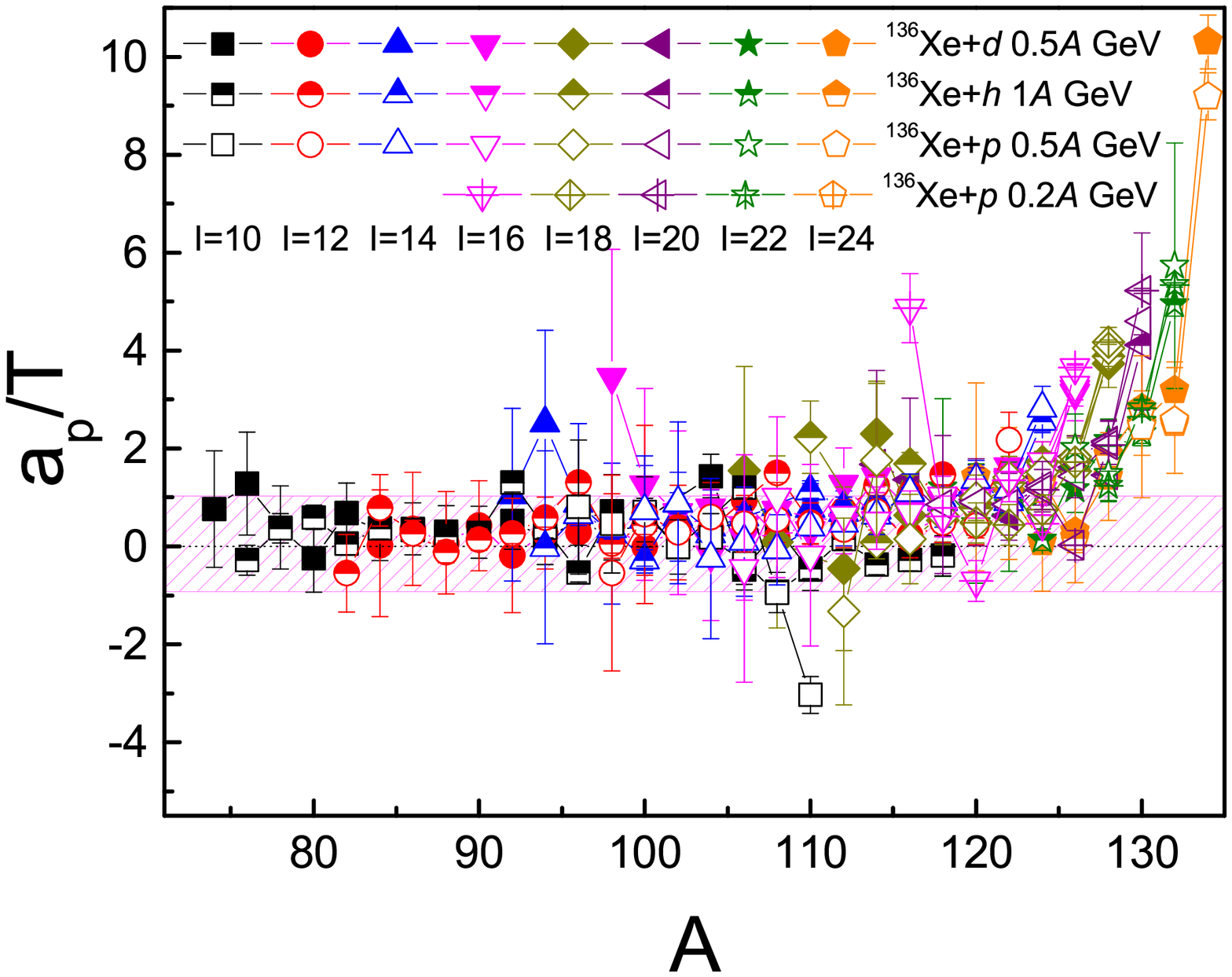}
\figcaption{\label{Xepdh} A comparison of $a_{p}/T$ for the measured fragments in the 0.2$A$, 0.5$A$, 1$A$ GeV $^{136}$Xe + $p$, and 0.5$A$ MeV $^{136}$Xe + $d$ spallation reactions. The shaded area denotes the range from -1 to 1. The cross sections for fragments in the 0.2$A$ GeV $^{136}$Xe + $p$ reaction are taken from Refs. \cite{Xep200}. The cross sections for fragments in the 0.5$A$ GeV $^{136}$Xe + $p$ and $^{136}$Xe + $d$ reactions are taken from Refs. \cite{Xep500} and \cite{Xed500}, and those for the 1$A$ GeV $^{136}$Xe + hydrogen ($p$) reaction are taken from Ref. \cite{Xep1000}.}
\end{center}

The fragments in the reactions analyzed above have neutron excesses of more than 20. In the 0.2$A$ GeV $^{136}$Xe +$p$ \cite{Xep200}, 0.5$A$ GeV $^{136}$Xe +$p$ \cite{Xep500}, $^{136}$Xe +$d$  \cite{Xed500}, and 1$A$ GeV $^{136}$Xe +$p$ \cite{Xep1000} spallation reactions, fragments with $I =$ 10 to 24 have been measured. The 0.2$A$ GeV $^{136}$Xe +$p$ spallation reaction was performed by C. Paradela \textit{et al} using the inverse kinematics technique at the FRS, GSI \cite{Xep200}, measuring the isotopes from  cadmium ($Z =$ 48) to cesium ($Z =$ 55). The 0.5$A$ MeV $^{136}$Xe +$p$ experiment was performed by L. Giot \textit{et al} using the FRS, GSI \cite{Xep500}, measuring the cross sections for isotopes from Nb ($Z=$ 41) to Ba ($Z =$ 56). The 0.5$A$ GeV $^{136}$Xe +$d$ experiment was performed by J. Alc\'{a}ntara-N\'{u}\~{n}ez \textit{et al} using the FRS, GSI \cite{Xed500}, measuring the cross sections for isotopes from V ($Z=$ 23) to Ba ($Z =$ 56). The 1$A$ GeV $^{136}$Xe + $p$ (hydrogen) reaction was performed by P. Napolitani \textit{et al} using the FRS, GSI \cite{Xep1000}, measuring the cross sections for isotopes from Li ($Z=$ 3) to Ba ($Z =$ 56). This provides an opportunity to study the evolution of $a_{p}/T$ of fragments in in-medium masses. The values of $a_{p}/T$ for most of the fragments with $A <$ 125 are within the range from -1 to 1. An increase of $a_{p}/T$ with $A$ is found in fragments with $A>$ 125, for which the mass numbers are quite close to that of $^{136}$Xe and are believed to be produced in peripheral collisions. Besides the 1$A$ GeV $^{136}$Xe + hydrogen ($p$) spallation reaction, the 1$A$ GeV $^{136}$Xe + $^{12}$C spallation reaction has also been measured (cross sections for isotopes not reported) using FRS and the large ALADIN dipole magnet \cite{Xe136C1,Xe136C2}. With more data provided, one can investigate the $^{136}$Xe spallation more systematically.

\begin{center}
\includegraphics[width=7.5cm]{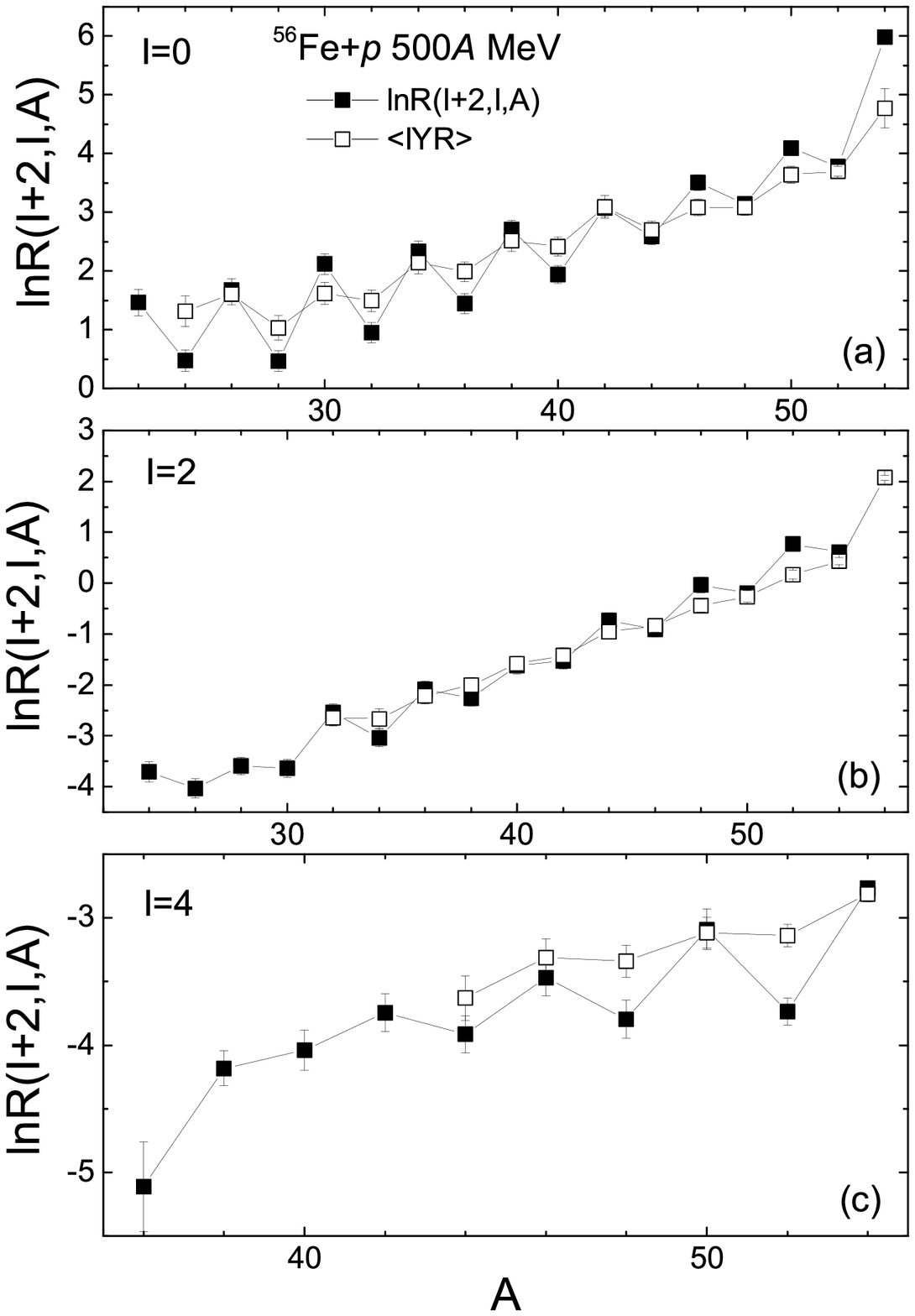}
\figcaption{\label{FepIYR} $\ln R(I+2,I,A)$ (solid symbols) and $<IYR>$ (open symbols) for the measured fragments in the 0.5$A$ GeV $^{56}$Fe + $p$ spallation reaction. The measured cross sections of fragments are taken from Ref. \cite{Fep56}.}
\end{center}

The smallest spallation system that can be found is the $^{56}$Fe + $p$ reactions, for which fragments were measured by C. Villagrase-Canton \textit{et al} at the FRS, GSI \cite{Fep56} at  incident energies of 0.3$A$, 0.5$A$, 0.75$A$, 1$A$, and 1.5$A$ GeV. The mass range of the measured fragments is from 20 to 54. The results of $\ln R(I+2,I,A)$ and $<IYR>$ for the measured fragments in 0.5$A$ GeV $^{56}$Fe + $p$ spallation reaction are plotted in Fig. \ref{FepIYR}. A strong odd-even staggering is found in the distribution of $\ln R(I+2, I, A)$ for the $I =$ 0 fragments. The $<IYR>$ reproduces the odd-even staggering, though the staggering is not as strong as that in the $\ln R(I+2,I,A)$.  The odd-even staggering is significantly weakened in $\ln R(I+2, I, A)$ for the $I =$ 2 fragments, and $<IYR>$ only has a small difference from $\ln R(I+2, I, A)$. To investigate the energy dependence of $a_{p}/T$, the results of $a_{p}/T$ for fragments are plotted in Fig. \ref{Fep315} for different values of $I$. A strong odd-even staggering is found in the values of $a_{p}/T$ for fragments with $I =$ 0 (a), while this staggering is weakened for fragments with $I =$ 2 (b) and 4 (c). The odd-even staggering phenomenon in $a_{p}/T$ has been also found in the fragments produced in projectile fragmentation reactions of 140$A$ MeV $^{40, 48}$Ca($^{58, 64}$Ni) + $^{9}$Be ($^{181}$Ta) \cite{Cw16}. A slight dependence of $a_{p}/T$ for fragments has been shown, in particular for the $I =$ 0 fragments. Huang \textit{et al} have attributed this phenomenon for light and symmetric fragments to the decay process of the last two or three particles, which happens at the end of deexcitation \cite{Mh10}, by a simulation using the antisymmetrized molecular dynamics plus {\sc gemini} method. The absolute values of $a_{p}/T$ for fragments are no larger than 4, and most of the values of $a_{p}/T$ for the fragments are within the range from -1 to 1.

\begin{center}
\includegraphics[width=8.cm]{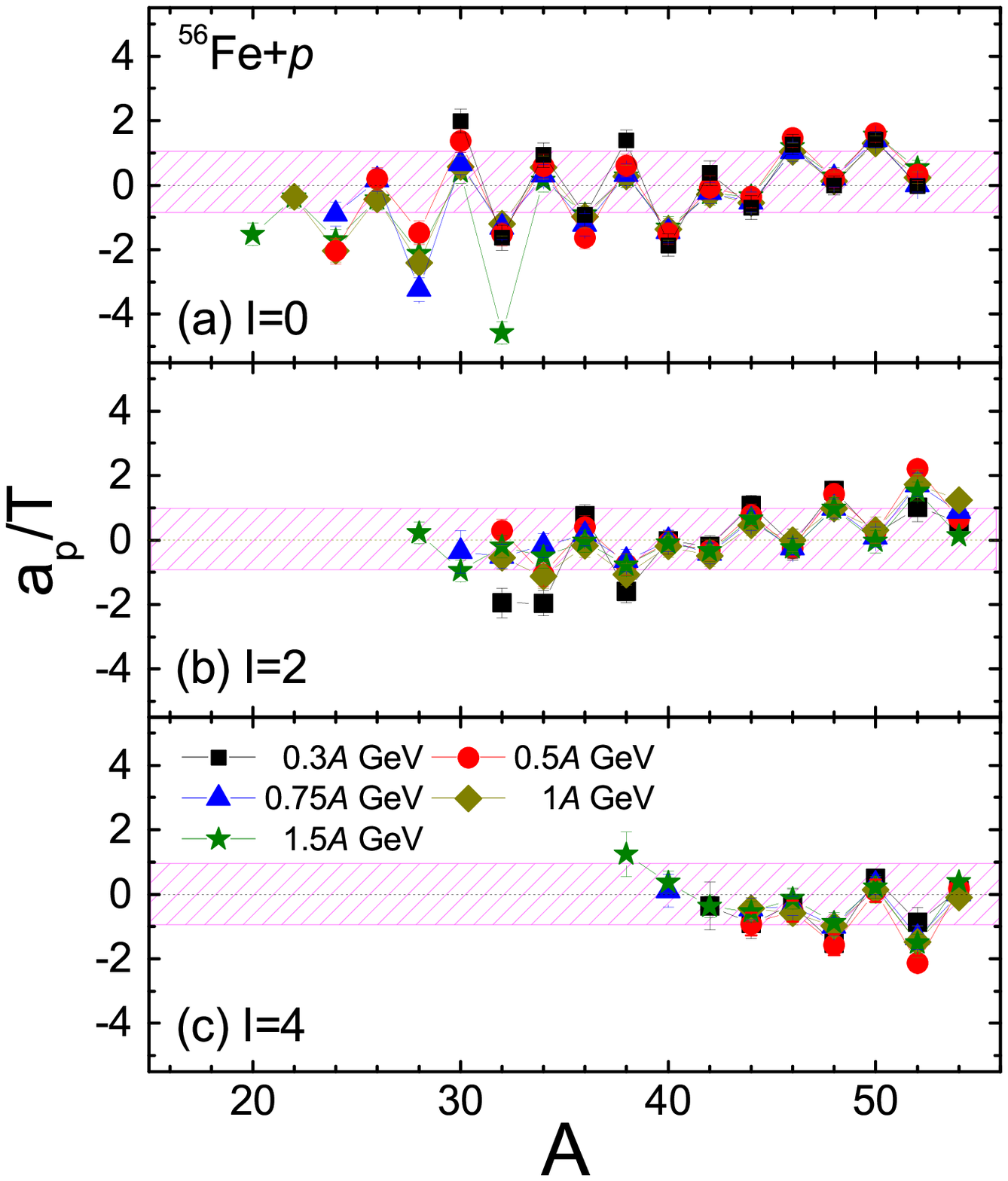}
\figcaption{\label{Fep315}  The values of $a_{p}/T$ for the measured fragment in the $^{56}$Fe + $p$ spallation reaction. The squares, circles, triangles, diamonds, and stars denote the results for fragments measured in the 0.3$A$, 0.5$A$, 0.75$A$, 1$A$, and 1.5$A$ GeV $^{56}$Fe+ $p$ reactions (measured cross sections of fragments taken from Ref. \cite{Fep56}), respectively. The results are plotted according to $I =$ 0 (a), 2 (b), and 4 (c), respectively. The shaded area in each panel denotes the range from -1 to 1.}
\end{center}

The results of $a_{p}/T$ for fragments in the $^{208}$Pb, $^{238}$U, $^{136}$Xe, and $^{56}$Fe spallation reactions have been found to be in a range from -4 to 4, while for most of the fragments the values are within a range from -1 to 1. The results agree with the $a_{p}/T$ obtained from the isotopic cross sections of high energy $p$ + Xe and $p$ + Kr reactions \cite{MFM84}, which are 1.80 and 1.63, respectively. As concluded in Ref. \cite{MFM84}, a reduced coefficient of the pairing energy term is expected in a system near its critical point. The  results obtained for $a_{p}/T$ should be correct for temperatures lower than the critical point ($\sim$ 5 MeV) as suggested in Ref. \cite{MFM84}.

An obvious odd-even staggering is shown in fragments with small $I$, and in relatively small-$A$ fragments which have large $I$. As has been explained in Ref. \cite{MRH10PRC}, the odd-even staggering is induced by the last few steps of particle emission in the de-excitation of hot fragments. The odd-even staggering phenomenon has also been investigated in the framework of the isospin-dependent quantum molecular dynamics (IQMD) model with the {\sc gemini} model by Su \textit{et al} \cite{SuJOE10}. Conclusions have been drawn that the odd-even staggering is affected by the excitation energies and the isotope distributions of the prefragments. The nuclear structure takes a important role in the formation of cold fragments. For  fragments with intermediate mass, where the nuclear structure is complex, the results for $a_{p}/T$ in this work may show a non-uniform staggering phenomenon. The phenomenon of staggering of $V(Z)$ (defined as $V(Z)=2\sigma(Z)/[\sigma(Z-1)+\sigma(Z+1)]$) was suggested as a puzzle in Ref. \cite{SuJOE10}.

The values of $a_{p}/T$ also depend on the mass number of the fragment, which is shown clearly for fragments with $A/A_{S} >$ 85\% ($A_{S}$ refers to the mass number of the spallation system). These fragments are mostly produced in semi-peripheral collisions which have relatively large impact parameters. In these zones, the large isospin difference induces a large difference between the densities of protons and neutrons. An enlarged odd-even staggering was also observed in the simulations of fragments yields in Ref.  \cite{SuJOE10}, which similarly reflects the dependence of the odd-even staggering strength on $A$, $Z$ and $T_{z}$ ($T_z = I/2$). The results of $a_p/t$ determined by the IYR method based on Eq. (\ref{apcal}) are sensitive to the IYR distributions for which $<IYR>$ is a good approximation for the symmetry energy and chemical potential energy terms for the even$-I$ fragments. The results shown in Fig. \ref{Pbp51000} indicate that the $<IYR>$ denoted by the odd-$I$ fragments are no longer good approximations for the symmetry energy and chemical potential terms for the even-$I$ fragments, where the large differences between $<IYR>$ and $\ln R(I+2, I, A)$ for the even-$I$ fragments are observed. A slight dependence of $a_{p}/T$ on the incident energy of the reaction is also observed.
In the form of pairing energy described in Eq. (\ref{aptcal}), the value of $a_{p} = $ 11.2 MeV is usually adopted. To estimate the cross section of a fragment, the value of $T$ is usually fixed at 2.0 MeV in the canonical ensemble theories \cite{GCE07}. A theoretical $a_{p}/T =$ 5.6 is suggested. The values of $a_{p}/T$ determined in this work, as well as the results reported for the fragments in the fragmentation reactions of $^{40, 48}$Ca ($^{58, 64}$Ni) \cite{Cw16}, are much smaller than 5.6. In Ref. \cite{Cw16}, it is explained as a temperature dependence of pairing energy with the help of a self-consistent finite-temperature relativistic Hartree-Bogoliubov model \cite{NYF13PRC}. In the de-excitation calculation of the statistical abrasion-ablation model \cite{ma14epja} and the SACA method \cite{SACA1,SACA2,SACA3,SACA4}, or the theoretical parameterizations of $T$-dependent binding energy using density-functional theory \cite{ParaTBE}, the temperature dependence of pairing energy is still not considered. In fact, the pairing energy of a fragment depends on mass in $1/A^{1/2}$ and the total pairing energy is small for fragments of large $A$, which suggests that the pairing energy may only have a small influence on the cross section of fragments. The results should be helpful in studying the properties of the very neutron-rich isotopes, and may help to study the sequential decay process of nuclear reactions.

\section{Summary}
\label{smmry}
The values of $a_{p}/T$ for fragments with $I \leq$ 36, which are measured in the $p$ + $^{56}$Fe/$^{136}$Xe/$^{208}$Pb/$^{238}$U spallation reactions, have been analyzed using an isobaric ratio method. Similar to the results for $a_{p}/T$ for fragments measured in projectile fragmentation reactions, the values of $a_{p}/T$ for neutron-rich fragments in spallation reactions fall into the range from -4 to 4 (most of them are within the range from -1 to 1), which agrees with the weakened pairing energy coefficients obtained in the early works on high-energy proton induced reactions on Xe and Kr. The $a_{p}/T$ of fragments with $A/A_{S} >$ 85\% are found to increase with $A$, which is explained by the approximation of symmetry energy and chemical potential terms by $<IYR>$ becoming worse. It is suggested that for cold fragments, which can be measured in experiment, a relatively small pairing energy should be adopted, especially when the spallation system is relatively small. The results should help in studying fragment production in spallation reactions for the ADS experiments, as well as in searching for isotopes with very large asymmetries.

\end{multicols}

\vspace{-1mm}
\centerline{\rule{80mm}{0.1pt}}
\vspace{2mm}

\begin{multicols}{2}

\end{multicols}

\clearpage
\end{CJK*}
\end{document}